\documentclass{PoS}
\usepackage{url}
\newcommand{\Ld}{\ensuremath{L_{\mathrm{D}}}}
\newcommand{\Lpl}{\ensuremath{L_{\mathrm{PL}}}}

\newcommand{\Lx}{\ensuremath{L_{\mathrm{X-rays}}}}
\newcommand{\Lopt}{\ensuremath{L_{\mathrm{Optical}}}}

\title{A link between radio loudness and X-ray/optical properties of AGN }

\ShortTitle{A link between radio loudness and X-ray/optical properties of AGN }

\author{\speaker{Sebastian Jester}%
         \thanks{Otto Hahn Fellow; next address: MPI f\"ur
           Astronomie, 69117 Heidelberg, Germany}, %
         Elmar K\"ording\thanks{Marie Curie intra-European Fellow}, %
         Rob Fender\\
        Department of Physics \& Astronomy, University of Southampton,
        Southampton SO17 1BJ, UK\\
        E-mail: \email{jester@phys.soton.ac.uk},
        \email{elmar@phys.soton.ac.uk}, \email{rpf@phys.soton.ac.uk}}

\abstract{We have found empirically that the radio loudness of AGN can
  be understood as function of both the X-ray and optical
  luminosity. This way of considering the radio loudness was inspired
  by the hardness-intensity diagrams for X-ray binaries, in which
  objects follow a definite track with changes to their radio
  properties occurring in certain regions.  We generalize the
  hardness-intensity diagram to a \emph{disk-fraction luminosity diagram},
  which can be used to classify the accretion states  both of X-ray
  binaries and of AGN.  Using a sample of nearly 5000 SDSS quasars with
  ROSAT matches, we show that an AGN is more likely to have a high
  radio:optical flux ratio when it has a high total luminosity or a
  large contribution from X-rays.  Thus, it is necessary to take into
  account both the optical and the X-ray properties of quasars in
  order to understand their radio loudness.  The success of
  categorizing quasars in the same way as X-ray binaries is further
  evidence for the unification of accretion onto stellar-mass and
  supermassive compact objects.}

\FullConference{VI Microquasar Workshop: Microquasars and Beyond\\
		 September 18-22, 2006\\
		 Como, Italy}

\begin{document}

\section{Introduction}
\label{s:intro}

The discovery of a common scaling relationship between radio
luminosity, X-ray luminosity and black-hole mass for Active Galactic
Nuclei (AGN) and Black-hole X-ray binaries (BHXRB) binaries
\cite{MHD03,FKM04} is a strong hint that accretion proceeds in a
similar way both for stellar-mass and supermassive black holes (see
also \cite{MKHea06,KFC06} and the contributions by K\"ording \& Fender
and Fender et al.\ in these proceedings).  A number of authors have
presented evidence that the accretion states that are observed in
BHXRB (see \cite{RM06} for a review) can also be observed in AGN (for
examples, see \cite{MGF03,MHD03,FKM04,Ho05,Jester05,GHU06}). Others
have searched for further scaling relationships and commonalities
between BHXRB and AGN, in particular with respect to timing properties
\cite{VU05,McHGUG05}, resulting in the recent discovery of a
fundamental timing plane shared by BHXRB and AGN \cite{McHKKea06}.

\begin{figure}
\hspace{0.25\hsize}\includegraphics[width=0.5\hsize]{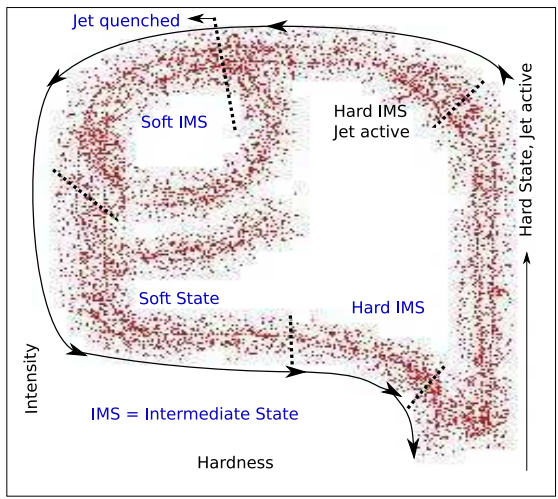}
\caption{\label{f:turtlehead}Sketch of the path traced by X-ray
  binaries in the hardness-intensity diagram (HID) during an outburst
  cycle.  The sketch shows the correlation between location of the
  object in the diagram and the jet properties (adapted from Fig.~7 in
  \protect\cite{FBG04}).}
\end{figure}
Recently, it has become clear that during an outburst cycle, BHXRBs
trace a characteristic path in a plot of X-ray hardness against
intensity (hardness-intensity diagram, HID) and that different regions
of the HID are associated with different radio properties
\cite{FBG04,BHCea05}.  Figure~\ref{f:turtlehead} gives a sketch of the
outburst cycle (sometimes called a ``turtlehead diagram'' because of
its characteristic shape resembling the animal, which has been called
mystic \cite{lu99}): the lowest-luminosity (low/hard) state is
associated with a continuous radio jet, which persists as the
luminosity increases and the object enters a hard intermediate state
(hard IMS).  At roughly constant luminosity, the object's spectrum then
softens to a soft IMS and radio observations show ``rapid ejections''
of material, probably indicating a transient but more highly
relativistic jet.  The spectrum then becomes entirely thermally
dominated (soft state) and simultaneously the radio emission is
quenched.  At lower accretion rates, the spectrum hardens again and
the jet radio emission reappears.

It has already been suggested that AGN with different radio properties
(radio-``loud'' vs.\ radio-``quiet'') correspond to BHXRB with
different accretion states \cite{MGF03,FKM04,Jester05,GHU06}.  Here,
we aim to put this analogy on a firm footing by constructing an
equivalent of HIDs that is physically meaningful for both BHXRB and
AGN, the disk-fraction luminosity diagram (DFLD).  The DFLD plots the
total luminosity (the sum of the thermal contribution from the
accretion disc, traced by the optical luminosity, and of non-thermal
emission, traced by the X-ray luminosity) against the fractional
contribution of non-thermal emission to the total. We find that in the
DFLD, AGN segregate by radio loudness, in a manner analogous to that
of BHXRBs in HIDs.  Thus, we find further evidence for the unification
of accretion in BHXRBs and AGN, and provide a framework to understand
the physical origin of radio loudness and quietness in AGN, which has
been a long-standing puzzle.  Full details of this work are published
in \cite{KJF06}, so this contribution will only highlight the salient
features of our approach and findings.

\section{Generalizing hardness-intensity to disk-fraction luminosity
  diagrams (DFLD)}
\label{s:theory}

In X-ray binaries, both the thermal emission from the accretion disc
and the non-thermal (power-law) X-ray emission (probably arising from
the disc's corona) are observed in the X-rays, as soft (peak at photon
energies $E_{\mathrm{peak}}\leq2.5$\,keV) and hard
($E_{\mathrm{peak}}\geq 50keV$) X-rays, respectively. The X-ray
hardness ratio measures the relative strength of the disc and
power-law emission, while the total intensity measures the total
accretion rate (although not necessarily in a linear way).  Because of
the scaling of disc temperature with black hole mass, $T \propto
M^{-1/4}$ \cite{SS73}, the disc emission in AGN is now seen in the
optical/UV wavelength region, while the power-law emission is still
seen in X-rays. Therefore, a pure X-ray hardness ratio misses the disc
emission in AGN and does not carry the same information as it does in
X-ray binaries.

Therefore, if we want to assess whether AGN behave in the same way as
BHXRB in terms of the HID, we need to generalize the HID in a
physically meaningful way.  As hardness equivalent, we use the
\emph{non-thermal fraction} $\Lpl/(\Ld+\Lpl)$, with the useful property
of being finite both for 0 and 100\% contribution from either
component, and set $\Ld=\Lopt$ and $\Lpl=\Lx$ for quasars (see
\cite{KJF06} for details on the computation of \Lopt\ and \Lx\ from
observed fluxes).  As intensity equivalent, we use the total
luminosity $\Ld+\Lpl$.  The advantage of our new definition is that it
can be used both for BHXRB and AGN.

\section{Constructing a DFLD for a sample of AGN}
\label{s:AGNsamp}

It was the aim of our work to consider whether AGN show the same
separation of accretion state and jet (radio) properties in DFLDs as
AGN.  In the case of X-ray binaries, individual objects can be
followed as they trace out their path in an DFLD during their outburst
cycles, which last of order days to months.  Since black-hole
timescales are expected to scale roughly linearly with mass, this is
not possible for AGN (although objects such as 3C\,120 \cite{MJGea02} might
be making small motions in the diagram, such as GRS1915+105 \cite{FBG04}).
Hence, we tackle our question in a statistical way, by considering the
average radio loudness of AGN as function of their location in the
DFLD.

\begin{figure}
\hspace{0.25\hsize}\includegraphics[clip=,angle=0,width=0.5\hsize]{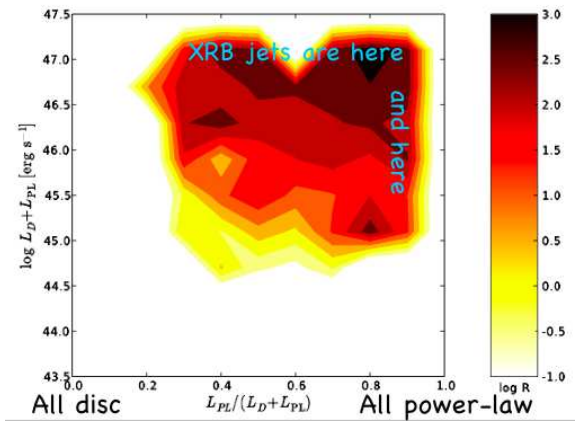}
\caption{\label{f:dfld_SDSS} Disc-fraction luminosity diagram (DFLD)
  showing average radio loudness of 4963 RASS-detected SDSS
  quasars. The contours give the average radio loudness
  $R=f(\mathrm{FIRST})/f(B)$ of the quasars in 10$\times$10 bins of
  total luminosity $\Ld+\Lpl$ and non-thermal fraction
  $\Lpl/(\Ld+\Lpl)$, with contour levels increasing by 0.5 dex.  The
  regions in which X-ray binaries have detectable radio emission from
  jets are also indicated.}
\end{figure}
\begin{figure}
\includegraphics[clip=,angle=0,width=\hsize]{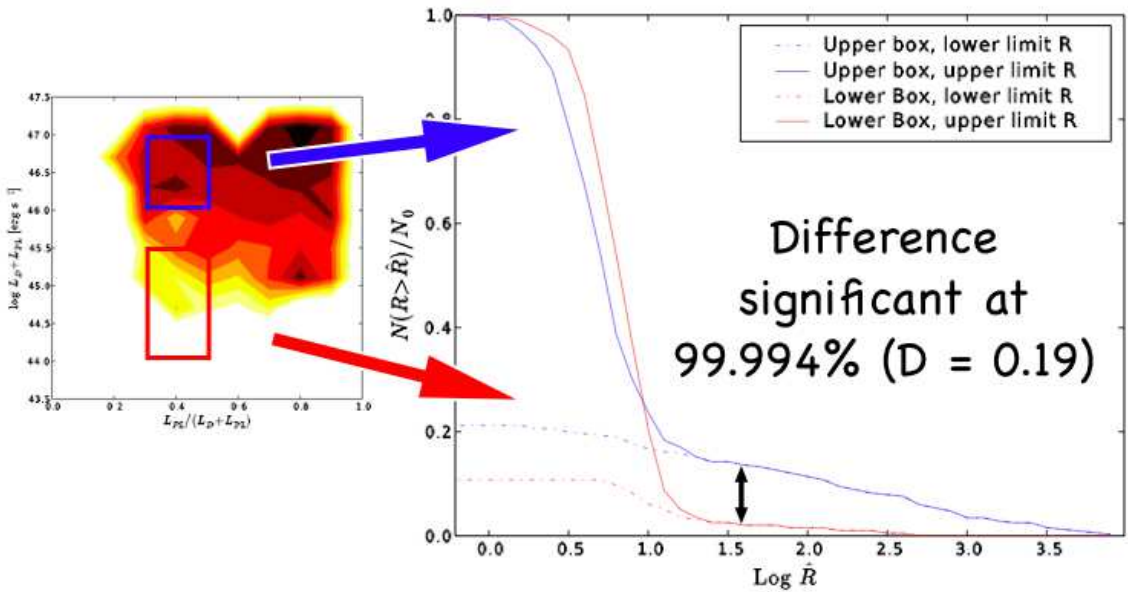}
\caption{\label{f:KStest}Comparison of the distributions of radio
  loudness in two areas of the DFLD.  A Kolmogorov-Smirnov test shows
  that there is a high statistical significance of the difference
  between the two radio loudness distributions.  This applies both to
  the radio loudness $R$ calculated by setting $R=0$ for sources
  undetected in FIRST (lower limit $R$), and for $R$ calculated by
  setting $R=R_{\mathrm{max}}$ (upper limit for $R$ from FIRST
  detection limit).}
\end{figure}
We use two AGN samples, beginning with objects from the 5th data
release (DR5) of the Sloan Digital Sky Survey (SDSS \cite{dr4}) that
are identified as quasar by the spectroscopic pipeline, a redshift in
the range $0.2 \leq z \leq 2.5$,  and that have
matches in the ROSAT All-Sky Survey (RASS).  Radio fluxes for this
sample are obtained from matches in the Faint Images of the Radio Sky
at Twenty centimeters (FIRST) survey. Both RASS and FIRST matches were
taken as provided in the SDSS Catalog Archive Server database
(CAS\footnote{\url{http://cas.sdss.org/astrodr5/}}).  There are a
total of 64268 objects classified as quasar in the DR5, of which 4963
have a RASS counterpart.
Figure~\ref{f:dfld_SDSS} shows the average radio loudness of the SDSS
quasars in 10$\times$10 bins of non-thermal fraction and total
luminosity.  Quasars tend to be more radio-loud on
average if they have a large total luminosity, and/or a large
non-thermal fraction.
\begin{figure}
\includegraphics[width=0.49\hsize]{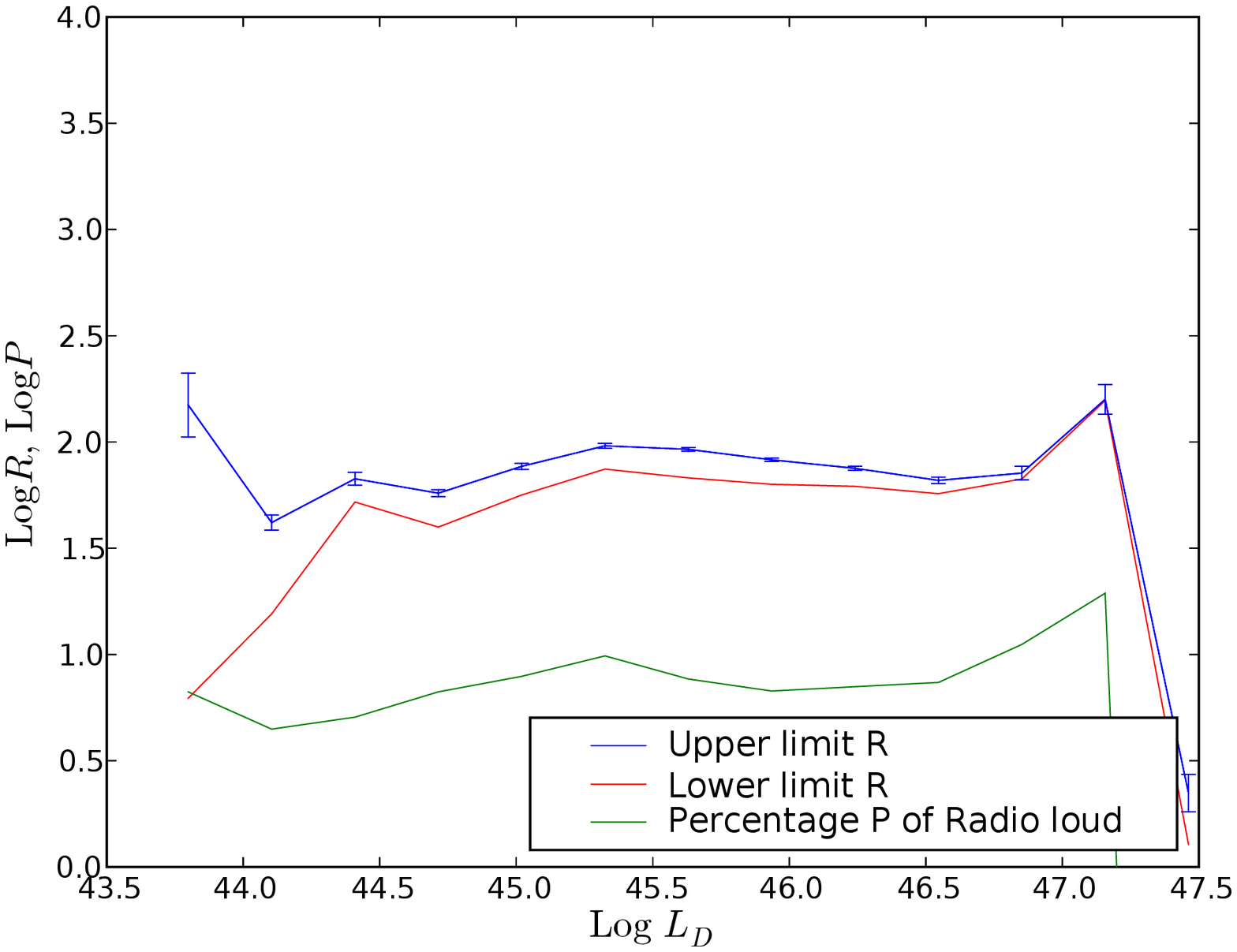}
\includegraphics[width=0.49\hsize]{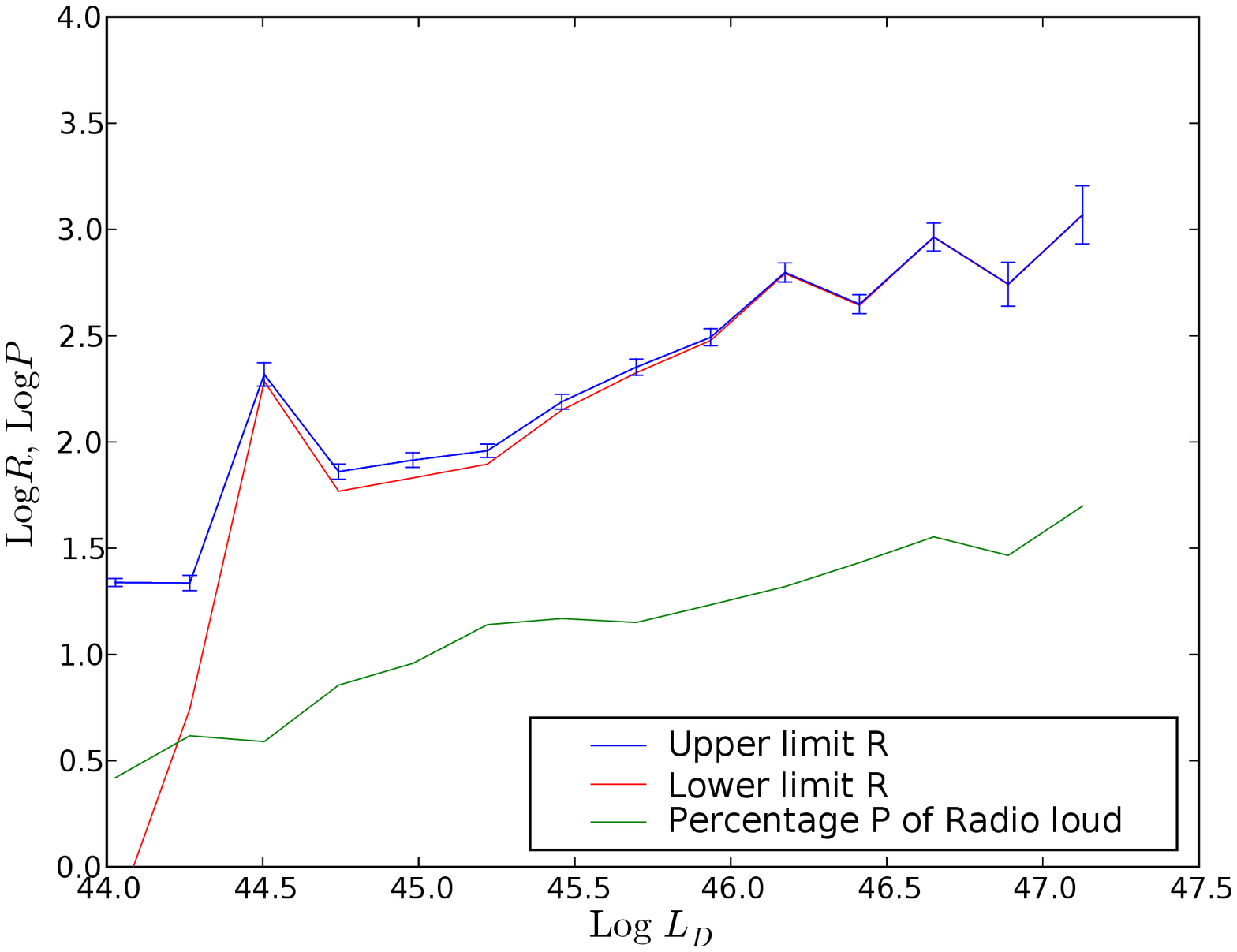}
\caption{\label{f:Rvsopt}Limits to average radio loudness and percentage of radio-loud
  sources ($R>10$) as function of disc luminosity. Left: all 64248
  SDSS quasars in the redshift range $0.2 \leq z \leq 2.5$. Right:
  only the 4963 X-ray detected sources.  The error bars give the
  uncertainty of the mean in each of the bins.  $R$ shows a
  strong correlation with optical luminosity when considering only the
 RASS-detected sources, but not when considering all quasars
 independently of their X-ray properties.}
\end{figure}
This difference is quantified by the Kolmogorov-Smirnov test shown in
Figure~\ref{f:KStest}. Thus, it is clear that the radio loudness of
AGN can only be understood by considering it as a function of
\emph{both} the optical and X-ray properties.  The relative
predominance of non-thermal emission reveals which accretion state the
disc is in, and this in turns governs the presence or absence of
powerful jets producing strong radio emission.

Figure~\ref{f:Rvsopt} gives another demonstration of the
interdependence of optical, X-ray and radio emission: when considering
the average radio loudness $R$ as function of the optical luminosity
for all SDSS quasars in our sample, there is no clear correlation with
the optical luminosity.  However, when the sample is restricted to
only the X-ray detected ones, the average $R$ clearly correlates with
optical luminosity.

Since the SDSS quasar survey is nearly exclusively sensitive to
luminous AGN, we now add a second sample of lower-luminosity
AGN. These are taken from \cite{HFS97} with X-ray fluxes taken from
the literature or archival data.  
\begin{figure}
\hspace{0.25\hsize}\includegraphics[angle=0,width=0.5\hsize]{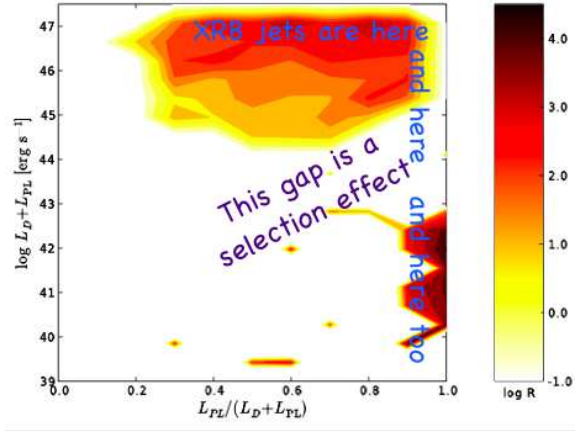}
\caption{\label{f:LLAGN}DFLD as in Fig.~\protect\ref{f:dfld_SDSS},
  with the addition of a sample of low-luminosity AGN (LLAGN).  The
  LLAGN lie in the area occupied by XRBs in the low/hard state, which
  show continuous jets.  They are much more radio-loud on average than
  the SDSS quasars.}
\end{figure}
In the DFLD, the LLAGN lie in the area occupied by XRBs in the
low/hard state (perhaps as expected from \cite{Ho99}), which show
continuous radio jets (see Figure~\ref{f:LLAGN}). The LLAGN have much
higher average $R$ values than the SDSS quasars, as found by
\cite{HP01}.

\section{Comparing DFLDs for AGN and XRBs}
\label{s:comp}

\begin{figure}
\includegraphics[width=0.49\hsize]{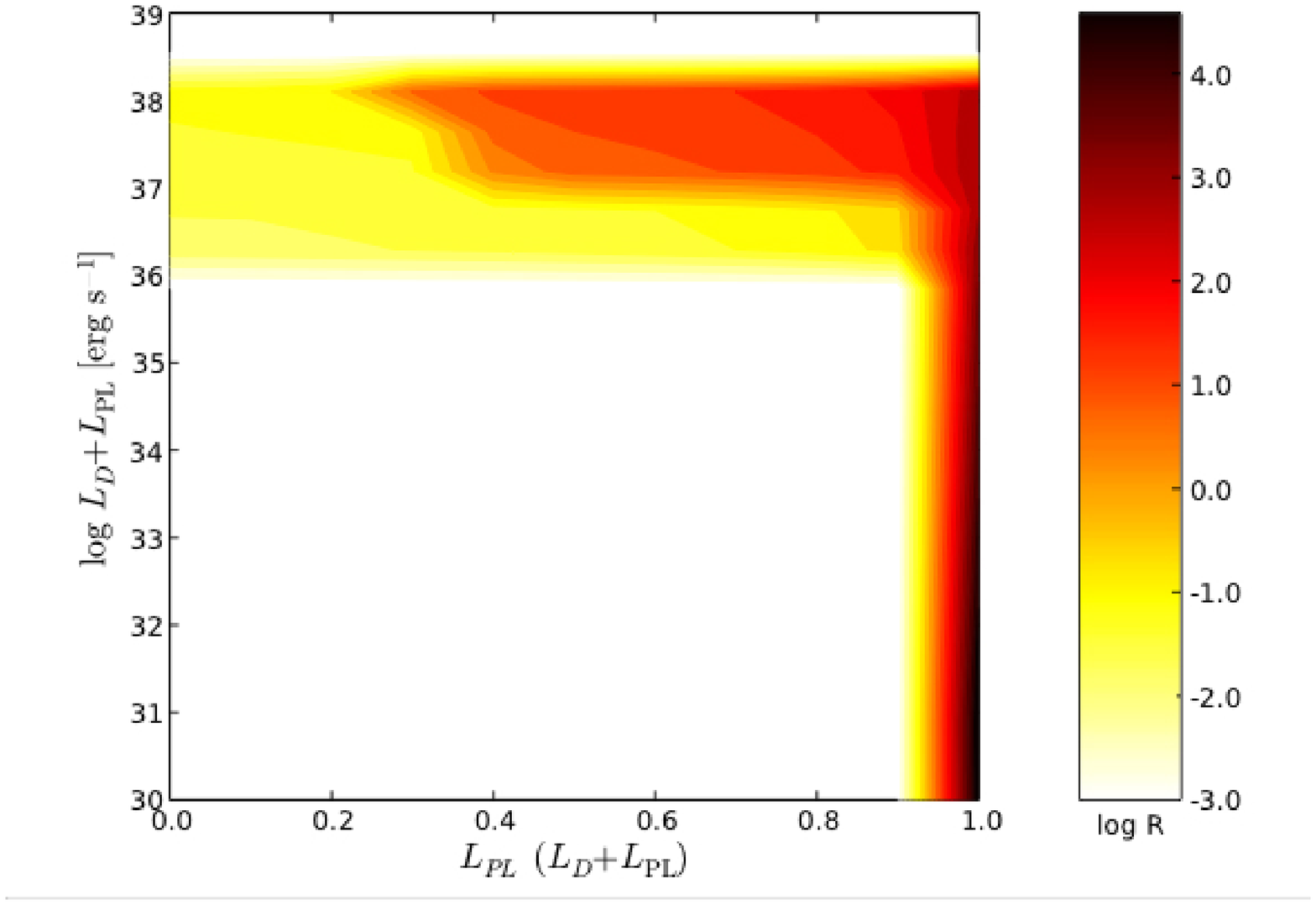}
\hfill
\includegraphics[width=0.49\hsize]{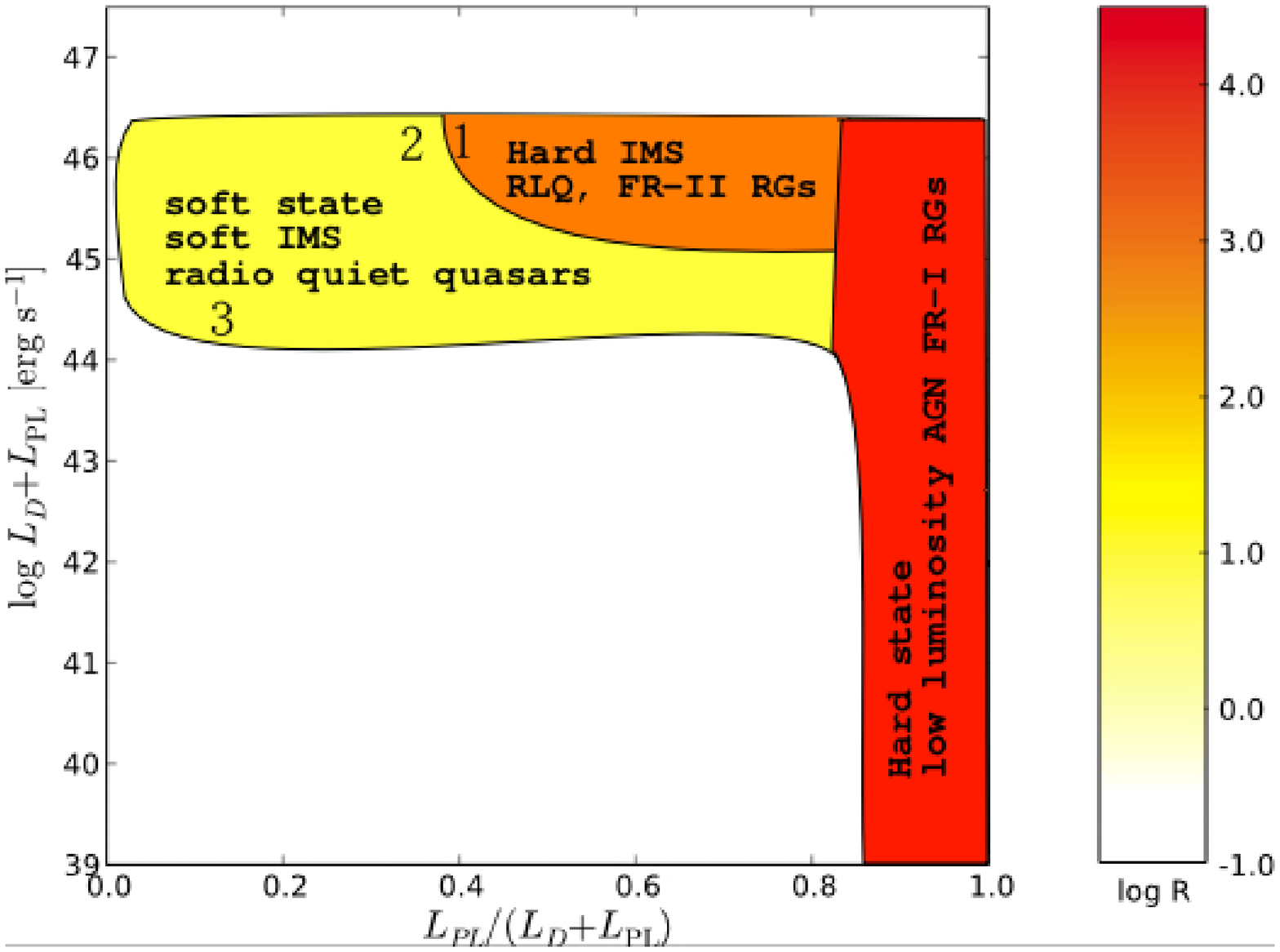}
\caption{\label{f:dfld_XRB}Left: DFLD for a simulated population of
  XRBs. Right: Unified picture of accretion states in XRBs and AGN.}
\end{figure}

So far, we have compared the DFLD for AGN and XRBs only qualitatively.
We would like to assess more quantitatively whether the population
DFLD for AGN, the only one we can construct within human timescales,
actually agrees with a population-averaged HID for XRBs.  However, the
observations (mostly by the \emph{Rossi X-ray Timing Explorer}) that
could be used to construct such an observed DFLD for a population of
XRBs are not readily accessible to us, and would suffer from
uncertainties in the distance, and hence luminosity, determination.
Therefore, we have performed a Monte-Carlo simulation of a population
of X-ray binaries in outburst cycles.  The left panel of
Figure~\ref{f:dfld_XRB} shows the resulting averaged DFLD.
The strong similarity between the population-averaged DFLD for XRBs
and that for our AGN sample strengthens the conclusion that there is a
universal accretion mechanism acting in discs both around stellar-mass
and around supermassive black holes, with similar disc-jet coupling in
both cases. This, like related earlier work
\cite{Mei01,Meier02,Meier03,MGF03,FKM04,KFC06}, suggests the identification
of AGN subclasses with XRB states shown in the right-hand panel of
Figure~\ref{f:dfld_XRB}.

\section{Outlook}
\label{s:outlook}

The presented work is only a first step towards establishing firmly
that AGN and XRBs have the same disc-jet coupling.  While we
considered many possible sources of error and selection effects, it
would be highly desirable to repeat our experiment with a sample that
has well-measured black-hole masses, and ideally with a volume-limited
sample, instead of the approximately flux-limited sample of SDSS
quasars used here.  In this context, it is a problem that
low-luminosity AGN have only been observed in the nearby Universe,
where only very few luminous quasars can be found. In XRBs, other
observable properties vary with HID position, in addition to the
presence or absence of jets: the timing properties \cite{BHCea05} and
(possibly) the jet Lorentz factor \cite{FBG04}.  If the picture
presented here is correct, equivalent correlations should be found in
AGN (compare \cite{VU05}).  Finally, even radio-``quiet'' quasars are
not completely radio-silent, implying that the jet is quenched but not
turned off completely.  The resulting prediction for XRBs is that even
objects in the thermally dominated soft state should show low-level
radio emission that could be found in very deep radio observations of
such objects.

\acknowledgments

SJ acknowledges support through an Otto Hahn Fellowship from the MPI
f\"ur Astronomie, Heidelberg. He is grateful to the members of the
Southampton astronomy group for their hospitality, and in particular
to Ian McHardy and Rob Fender (as well as PPARC) for infrastructure
and travel support, and to the RAS for a travel grant to Como.


\begin{thebibliography}{99}


\bibitem{dr4}
J.~K. {Adelman-McCarthy} et al., \emph{{The Fourth Data Release of the Sloan Digital Sky Survey}},
  \apjs\, \textbf{162}, 38--48 (2006) [{\tt astro-ph/0507711}].

\bibitem{BHCea05}
T.~{Belloni}, J.~{Homan}, P.~{Casella}, M.~{van der Klis}, E.~{Nespoli},
  W.~H.~G. {Lewin}, J.~M. {Miller}, and M.~{M{\'e}ndez}, \emph{{The evolution
  of the timing properties of the black-hole transient GX 339-4 during its
  2002/2003 outburst}}, \aap\, \textbf{440}, 207--222 (2005) [{\tt
  astro-ph/0504577}].

\bibitem{FKM04}
H.~{Falcke}, E.~{K{\"o}rding}, E., and S.~{Markoff}, \emph{{A scheme to unify
  low-power accreting black holes. Jet-dominated accretion flows and the
  radio/X-ray correlation}}, \aap\, \textbf{414}, 895--903 (2004).

\bibitem{FBG04}
R.~P. {Fender}, T.~M. {Belloni}, and E.~{Gallo}, \emph{{Towards a unified model
  for black hole X-ray binary jets}}, \mnras\, \textbf{355}, 1105--1118 (2004).

\bibitem{GHU06}
J.~E. {Greene}, L.~C. {Ho}, and J.~S. {Ulvestad}, \emph{{The Radio Quiescence
  of Active Galaxies with High Accretion Rates}}, \apj\, \textbf{636}, 56--62
  (2006) [{\tt astro-ph/0509198}].

\bibitem{Ho99}
L.~C. {Ho}, \emph{{The Spectral Energy Distributions of Low-Luminosity Active
  Galactic Nuclei}}, \apj\, \textbf{516}, 672--682 (1999).

\bibitem{Ho05}
L.~C. {Ho}, \emph{{``Low-State'' Black Hole Accretion in Nearby Galaxies}},
  \apss\, \textbf{300}, 219--225 (2005) [{\tt astro-ph/0504643}].

\bibitem{HFS97}
L.~C. {Ho}, A.~V. {Filippenko}, and W.~L.~W. {Sargent}, \emph{{A Search for
  ``Dwarf'' Seyfert Nuclei. III. Spectroscopic Parameters and Properties of the
  Host Galaxies}}, \apjs\, \textbf{112}, 315--+ (1997) [{\tt
  astro-ph/9704107}].

\bibitem{HP01}
L.~C. {Ho} and C.~Y. {Peng}, \emph{{Nuclear Luminosities and Radio Loudness of
  Seyfert Nuclei}}, \apj\, \textbf{555}, 650--662 (2001).

\bibitem{Jester05}
S.~{Jester}, \emph{{A Simple Test for the Existence of Two Accretion Modes in
  Active Galactic Nuclei}}, \apj\, \textbf{625}, 667--679 (2005) [{\tt
  astro-ph/0502394}].

\bibitem{KFC06}
E.~{K{\"o}rding}, H.~{Falcke}, and S.~{Corbel}, \emph{{Refining the fundamental
  plane of accreting black holes}}, \aap\, \textbf{456}, 439--450 (2006) [{\tt
  astro-ph/0603117}].

\bibitem{KJF06}
E.~G. {K{\"o}rding}, S.~{Jester}, and R.~{Fender}, \emph{{Accretion states and
  radio loudness in active galactic nuclei: analogies with X-ray binaries}},
  \mnras\, \textbf{in press} (2006) [{\tt astro-ph/0608628}].

\bibitem{lu99}
John {Lasseter} and Lee {Unkrich}, \emph{Toy story 2}, 1999.

\bibitem{MGF03}
T.~J. {Maccarone}, E.~{Gallo}, and R.~{Fender}, \emph{{The connection between
  radio-quiet active galactic nuclei and the high/soft state of X-ray
  binaries}}, \mnras\, \textbf{345}, L19--L24 (2003).

\bibitem{MJGea02}
A.~P. {Marscher}, S.~G. {Jorstad}, J.~{G{\'o}mez}, M.~F. {Aller},
  H.~{Ter{\"a}sranta}, M.~L. {Lister}, and A.~M. {Stirling},
  \emph{{Observational evidence for the accretion-disk origin for a radio jet
  in an active galaxy}}, \nat\, \textbf{417}, 625--627 (2002).

\bibitem{McHGUG05}
I.~M. {McHardy}, K.~F. {Gunn}, P.~{Uttley}, and M.~R. {Goad},
  \emph{{MCG-6-30-15: long time-scale X-ray variability, black hole mass and
  active galactic nuclei high states}}, \mnras\, \textbf{359}, 1469--1480
  (2005) [{\tt astro-ph/0503100}].

\bibitem{McHKKea06}
I.~M. {McHardy}, E.~{K{\"o}rding}, C.~{Knigge}, P.~{Uttley}, and R.~P.
  {Fender}, \emph{Active galactic nuclei as galactic black holes scaled in both
  mass and accretion rate}, \nat\, \textbf{in press} (2006).

\bibitem{Mei01}
D.~L. {Meier}, \emph{{The Association of Jet Production with Geometrically
  Thick Accretion Flows and Black Hole Rotation}}, \apjl\, \textbf{548},
  L9--L12 (2001).

\bibitem{Meier02}
D.~L. {Meier}, \emph{{Grand unification of AGN and the accretion and spin
  paradigms}}, New Astronomy Review\, \textbf{46}, 247--255 (2002).

\bibitem{Meier03}
D.~L. {Meier}, \emph{{The theory and simulation of relativistic jet formation:
  towards a unified model for micro- and macroquasars}}, New Astronomy Review\,
  \textbf{47}, 667--672 (2003).

\bibitem{MHD03}
A.~{Merloni}, S.~{Heinz}, and T.~{di Matteo}, \emph{{A Fundamental Plane of
  black hole activity}}, \mnras\, \textbf{345}, 1057--1076 (2003).

\bibitem{MKHea06}
A.~{Merloni}, E.~{K{\"o}rding}, S.~{Heinz}, S.~{Markoff}, T.~{Di Matteo}, and
  H.~{Falcke}, \emph{{Why the fundamental plane of black hole activity is not
  simply a distance driven artifact}}, New Astronomy\, \textbf{11}, 567--576
  (2006) [{\tt astro-ph/0601286}].

\bibitem{RM06}
R.~A. {Remillard} and J.~E. {McClintock}, \emph{{X-Ray Properties of Black-Hole
  Binaries}}, \araa\, \textbf{44}, 49--92 (2006) [{\tt astro-ph/0606352}].

\bibitem{SS73}
N.~I. {Shakura} and R.~A. {Sunyaev}, \emph{{Black holes in binary systems.
  Observational appearance.}}, \aap\, \textbf{24}, 337--355 (1973).

\bibitem{VU05}
S.~{Vaughan} and P.~{Uttley}, \emph{{Where are the X-ray quasi-periodic
  oscillations in active galaxies?}}, \mnras\, \textbf{362}, 235--244 (2005)
  [{\tt astro-ph/0506455}].

\end{thebibliography}

\end{document}